\documentclass[12pt,preprint]{aastex}

\begin{document}
\slugcomment{Accepted to ApJ Letters 2008 April 17}
\title{Dead Zone Accretion Flows in Protostellar Disks}

\author{N. J. Turner\altaffilmark{1} and T. Sano\altaffilmark{2}}

  \altaffiltext{1}{Jet Propulsion Laboratory, California Institute of
    Technology, Pasadena, California 91109, USA;
    neal.turner@jpl.nasa.gov}

  \altaffiltext{2}{Institute of Laser Engineering, Osaka University,
    Suita, Osaka 565-0871, Japan; sano@ile.osaka-u.ac.jp}

\begin{abstract}
  Planets form inside protostellar disks in a dead zone where the
  electrical resistivity of the gas is too high for magnetic forces to
  drive turbulence.  We show that much of the dead zone nevertheless
  is active and flows toward the star while smooth, large-scale
  magnetic fields transfer the orbital angular momentum radially
  outward.  Stellar X-ray and radionuclide ionization sustain a weak
  coupling of the dead zone gas to the magnetic fields, despite the
  rapid recombination of free charges on dust grains.  Net radial
  magnetic fields are generated in the magneto-rotational turbulence
  in the electrically conducting top and bottom surface layers of the
  disk, and reach the midplane by Ohmic diffusion.  A toroidal
  component to the fields is produced near the midplane by the orbital
  shear.  The process is similar to the magnetization of the Solar
  tachocline.  The result is a laminar, magnetically-driven accretion
  flow in the region where the planets form.
\end{abstract}


\section{INTRODUCTION}

The planets of Sun-like stars form from a protostellar disk consisting
of 0.01--0.1~Solar masses of gas together with 1\% dust by mass.  The
removal of orbital angular momentum from the disk is a key to
understanding the origins of the planets as it determines the rate at
which material spirals in to accrete on the star.  The
magneto-rotational instability or MRI \citep{bh91} transfers angular
momentum outward and taps the free energy in the differential orbital
rotation of the disk to drive turbulence and regenerate magnetic
fields, at locations where the gas is sufficiently ionized to couple
strongly to the fields.  However the minimum-mass protosolar disk,
constructed by adding sufficient hydrogen and helium to the planets to
give Solar composition \citep{hn85}, is mostly too cold for thermal
ionization.  The minimum-mass disk has surface mass density
$\Sigma=1700(r/$AU$)^{-3/2}$~g~cm$^{-2}$ and temperature
$T=280(r/$AU$)^{-1/2}$~K at radius $r$ measured in astronomical units
(AU), and is in vertical hydrostatic balance.  Cosmic rays and stellar
coronal X-rays are absorbed before reaching the midplane, and ionize
only the surface layers.  Furthermore, rapid recombination on grain
surfaces leads to low abundances of the free electrons otherwise
responsible for most of the conductivity.  Consequently, much of the
region where the planets form is not subject to the MRI.  The region
consists of a laminar resistive interior sandwiched between two
conducting surface layers where magneto-rotational turbulence causes
an accretion flow \citep{g96,sm00,ft02,in06a}.  The interior is not
completely quiescent, as waves propagating from the turbulent layers
cause hydrodynamic stresses and low rates of accretion \citep{fs03}.
Magnetic stresses can occur in the interior under favorable
conditions, if small grains are removed, for example by incorporation
into planetesimals, and ionizing cosmic rays reach the disk unimpeded
by the stellar wind \citep{ts07}.  However, the resistive interior of
the disk retains a weak coupling to the magnetic fields even if grains
are present and cosmic rays are absent.  The weak coupling is the
topic of this paper.

\section{CRITERION FOR SHEAR-GENERATED MAGNETIC FIELDS}

Magnetic forces can extract angular momentum from the gas if the
fields have both toroidal and poloidal components.  We derive a
criterion for the generation of toroidal fields by shear acting on
radial fields.  The Ohmic diffusion quickly eliminates any vertical
magnetic gradients in the resistive layer, while the conducting layers
above and below prevent the fields from escaping.  The resistive layer
can be approximated by an axisymmetric cylindrical flow in Keplerian
rotation with a uniform resistivity $\eta$.  The toroidal component of
the induction equation reduces to $\partial B_\phi/\partial t =
-\frac{3}{2}\Omega B_r + \eta(\nabla^2B_\phi-B_\phi/r^2)$ where $B_r$
and $B_\phi$ are the radial and toroidal field components in
cylindrical coordinates and $\Omega$ is the orbital frequency.  If
also the toroidal field varies as a power-law with radius, $B_\phi\sim
r^p$, the field grows through the shear term provided $r^2\Omega/\eta
> \frac{2}{3}(p^2-1)B_\phi/B_r$.  We consider the implications for the
minimum-mass protosolar disk.  The magnetic pressure is unlikely to
exceed the midplane gas pressure, owing to the buoyancy of the fields.
The magnetic pressure is a fixed fraction of the midplane gas pressure
if the power-law index $p=-13/8$.  The resulting criterion for
shear-generated toroidal fields to grow to about ten times the radial
seed field strength,
\begin{equation}\label{eqn:shear}
  v_\phi^2/(\eta\Omega)>10,
\end{equation}
resembles the criterion \citep{ss02b} for magneto-rotational
turbulence $v_{Az}^2/(\eta\Omega)>1$.  However the shear criterion
equation~\ref{eqn:shear} allows magnetic activity at far greater
resistivities, since the flattened shape of the disk requires an
orbital speed $v_\phi$ much greater than the sound speed $c_s$, which
in turn is typically greater than the vertical Alfv\'en speed
$v_{Az}$.  The orbital speed is a few hundred times the vertical
Alfv\'en speeds found at the midplane in numerical calculations of
magneto-rotational turbulence \citep{ms00}.  The part of the dead zone
where the shear criterion holds is better named the ``undead zone''
because it can be reactivated by radial fields siphoned off from the
adjacent turbulent layers.  In contrast with models where angular
momentum is removed from the top and bottom surfaces of the disk by a
magneto-centrifugal wind \citep{bp82,wk93}, here the torques arise
from locally-generated magnetic fields.

\section{RESISTIVITY OF THE PROTOSOLAR DISK}

\begin{figure}[tb!]
  \epsscale{0.75}
  \plotone{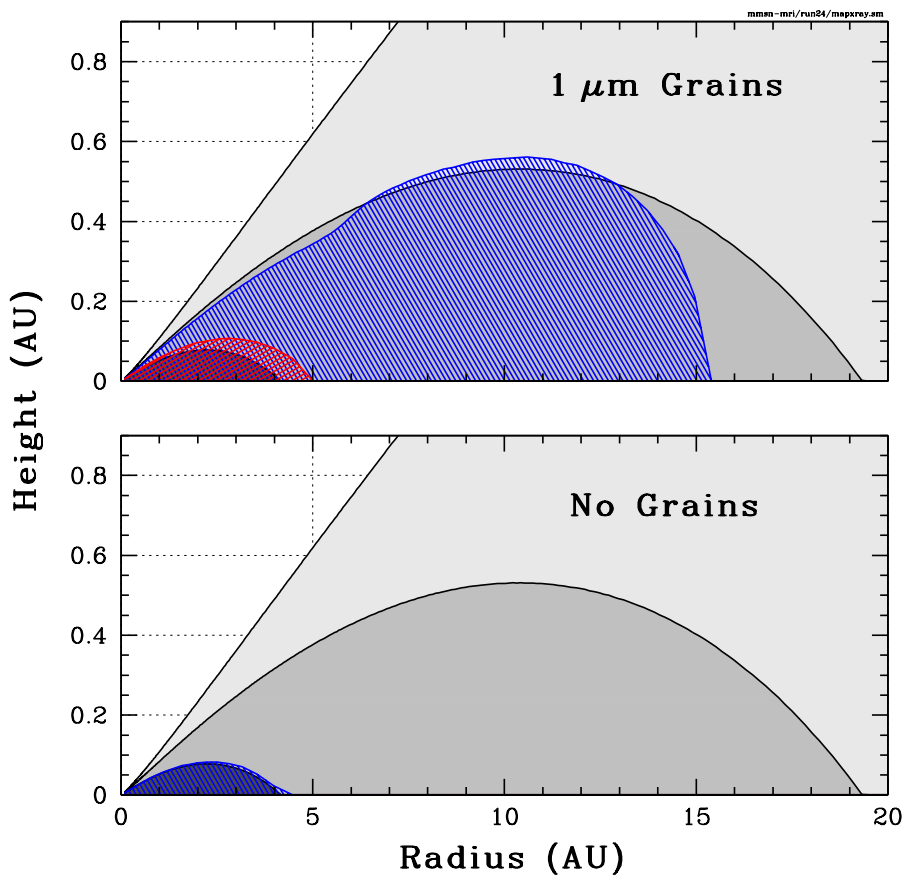}

  \figcaption{\sf Location of the dead zone (red) and undead zone
    (blue) in a cross-section view of the minimum-mass protosolar
    disk.  The dead zone gas is decoupled from the magnetic fields,
    while the undead zone is sufficiently ionized for the generation
    of toroidal fields by shear.  The remainder of the disk body is
    well-coupled to the fields and is turbulent through
    magnetorotational instability.  The star lies at the origin, the
    disk midplane falls along the horizontal axis, and solid black
    contours show vertical mass columns of 1, 10 and 100~g~cm$^{-2}$.
    The pressure in the vertical component of the magnetic field is
    set to 0.1\% of the midplane gas pressure.  The ionization is due
    to stellar X-rays and the decay of radioactive $^{26}$Al.
    Recombination on grain surfaces is included in the calculation
    shown at top, where the dead zone extends to 5~AU in the midplane.
    The grains are 1~$\mu$m in radius and well-mixed in the gas at a
    1\% mass fraction.  With the grains removed, there is no dead zone
    (bottom panel).\label{fig:mapxray}}
\end{figure}

The minimum-mass protosolar disk with well-mixed micron-sized grains
has an equilibrium resistivity low enough for magneto-rotational
turbulence only in the surface layers, at the radii where most of the
planets formed (figure~\ref{fig:mapxray}).  A deeper layer can
generate smooth large-scale fields by shear according to
equation~\ref{eqn:shear}, while the midplane within 5~AU of the star
is almost completely decoupled from the fields.  If the grains are
removed by incorporation into planetary embryos, then turbulence or
shear can generate magnetic fields in every part of the disk.

Figure~\ref{fig:mapxray} was constructed as follows.  The resistivity
$\eta = 234\sqrt{T}/x_e$~cm$^2$~s$^{-1}$ varies inversely with the
electron fraction $x_e=n_e/n_n$, where $n_e$ is the electron number
density and $n_n$ the number density of neutrals \citep{bb94}.  We
compute the electron fraction by solving a chemical network including
ionization by stellar X-rays \citep{ig99} and $^{26}$Al radionuclide
decay, dissociative recombination, charge exchange with metal atoms,
and grain surface reactions.  The recombination is treated using the
reduced reaction network with grains described by \cite{in06a}.  The
fraction of the metal atoms free to enter the gas phase is chosen to
be 1\%.  The resistivity depends only weakly on this fraction.  The
stellar X-ray ionization rate is taken from a fit to Monte Carlo
radiative transfer results including scattering \citep{ig99}, scaled
to a stellar luminosity $2\times 10^{30}$~erg~s$^{-1}$ in 5~keV
thermal X-rays to match young Solar-mass stars observed in the Orion
nebula \citep{gf00}.  The fitted ionization rate is $\zeta=2.6\times
10^{-15}(r/$AU$)^{-2} (\exp[-\Sigma_a/8.0$~g~cm$^{-2}] +
\exp[-\Sigma_b/8.0$~g~cm$^{-2}])$~s$^{-1}$ where $\Sigma_a$ and
$\Sigma_b$ are the mass columns lying vertically above and below the
point of interest.  The fit underestimates the ionization rate at mass
columns $<1$~g~cm$^{-2}$, with no significant effect on the locations
of the dead and undead zones.  The decay of $^{26}$Al in the dust
grains yields an ionization rate $4\times 10^{-19}$~s$^{-1}$
\citep{s92}.  A fraction 0.1\% of the radioactive atoms is placed in
the gas phase, giving a low rate of ionization even with grains
absent.  We neglect cosmic ray ionization owing to uncertainty about
the extent to which the interstellar energetic particles are blocked
by the wind from the young star.

\section{MHD CALCULATIONS}

We show results from two 3-D isothermal stratified shearing-box
\citep{hg95,bn95,sh96} MHD calculations of small patches of the
minimum-mass protosolar disk using the ZEUS code \citep{sn92b}.  The
first is placed at 5~AU and includes well-mixed 1~$\mu$m grains with
dust-to-gas mass ratio 1\% and a magnetic field with a net vertical
component of 6~milligauss.  The second is placed at 1~AU and has no
grains and a net vertical field of 30~milligauss.  The net magnetic
fields are weak, with midplane ratios of gas to magnetic pressure
$5\times 10^4$ and $3.8\times 10^5$, respectively.  The presence of
the net fields ensures that the fastest-growing mode of the linear MRI
is spatially resolved in the calculations.  To the net field in both
cases is added a part with zero net flux that dissipates readily
through reconnection.  The vertical component is $B_0\sin 2\pi x/L$,
the toroidal component $B_0\cos 2\pi x/L$ and the radial component
zero, where $x$ is the radial position and $L$ the box width.  The
strength $B_0$ of the sinusoidal field is 0.1 and 1~Gauss,
respectively.  The resistivity is calculated using the ionization and
chemical network described above except that the X-ray ionization
rates are directly interpolated in the Monte Carlo radiative transfer
results and the fit discussed above is not used.  The resistivity is
fixed at its initial equilibrium value at each height in the first
run, while the resistivity is allowed to vary in space and time
through mixing, ionization and recombination in the second run.  In
other respects the calculations are identical to our recent work
\citep{ts07}.  In particular, the vertical boundaries lie four density
scale heights $H=c_s/\Omega$ above and below the midplane and are
open, allowing magnetic fields to escape leaving the patch of disk
with a net radial and toroidal magnetic flux.  Among the non-ideal
terms in the induction equation, ambipolar diffusion is important
outside $4H$.  The Ohmic resistivity is greater than the Hall
resistivity near the midplane at 1~AU while the two are comparable at
5~AU \citep{w07}.  Under these conditions the Ohmic term has the
bigger impact on the amplitude of magneto-rotational turbulence
\citep{ss02b}.  Our calculations therefore include only the Ohmic
diffusion.  The domain extends $2H$ along the radial direction and
$8H$ along the toroidal direction, and is divided into 32 radial by 64
toroidal by 128 vertical zones.

\begin{figure}[tb!]
  \epsscale{0.75}
  \plotone{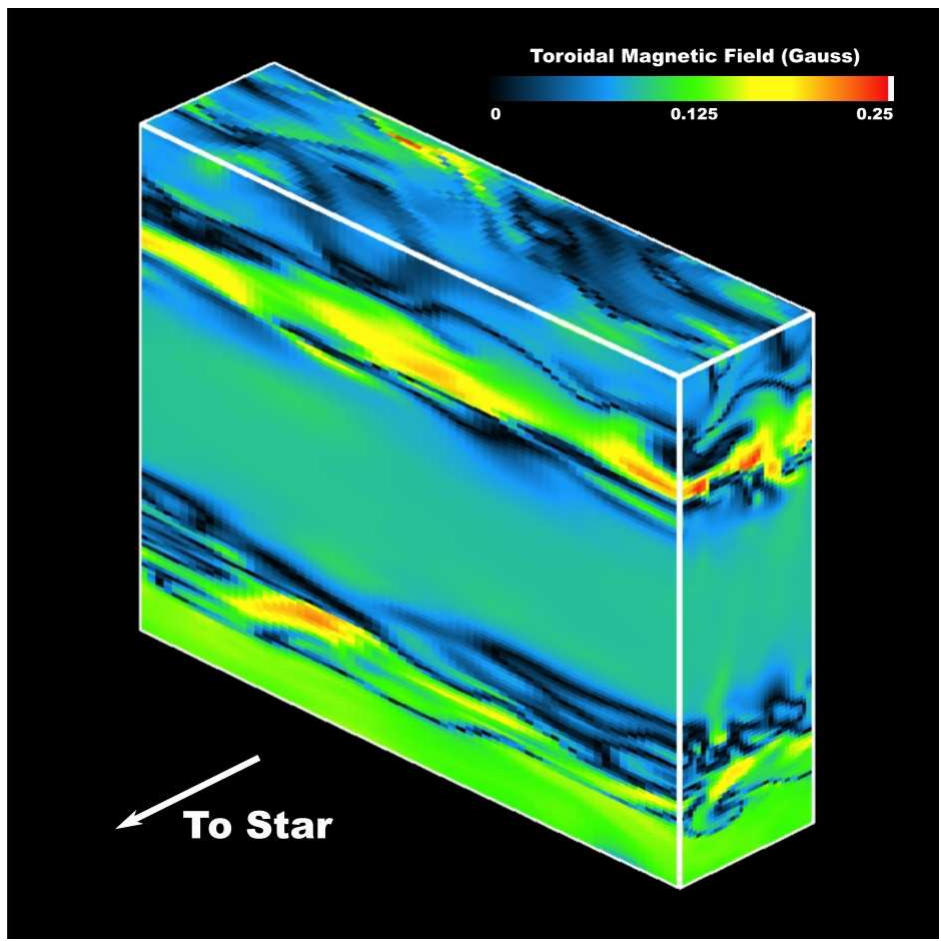}

  \figcaption{\sf Snapshot of the toroidal magnetic field strength at
    55~orbits in a resistive MHD calculation of a patch of the
    protosolar disk at 5~AU including well-mixed 1~$\mu$m grains.  The
    undead zone at center is filled with a uniform, 0.1-Gauss
    shear-generated toroidal magnetic field while patchy fields are
    found in the turbulent layers above and below.  The star lies
    off-page to lower left and the disk midplane is horizontal through
    the image center.\label{fig:snapshot}}
\end{figure}

\begin{figure}[tb!]
  \epsscale{1.0}
  \plotone{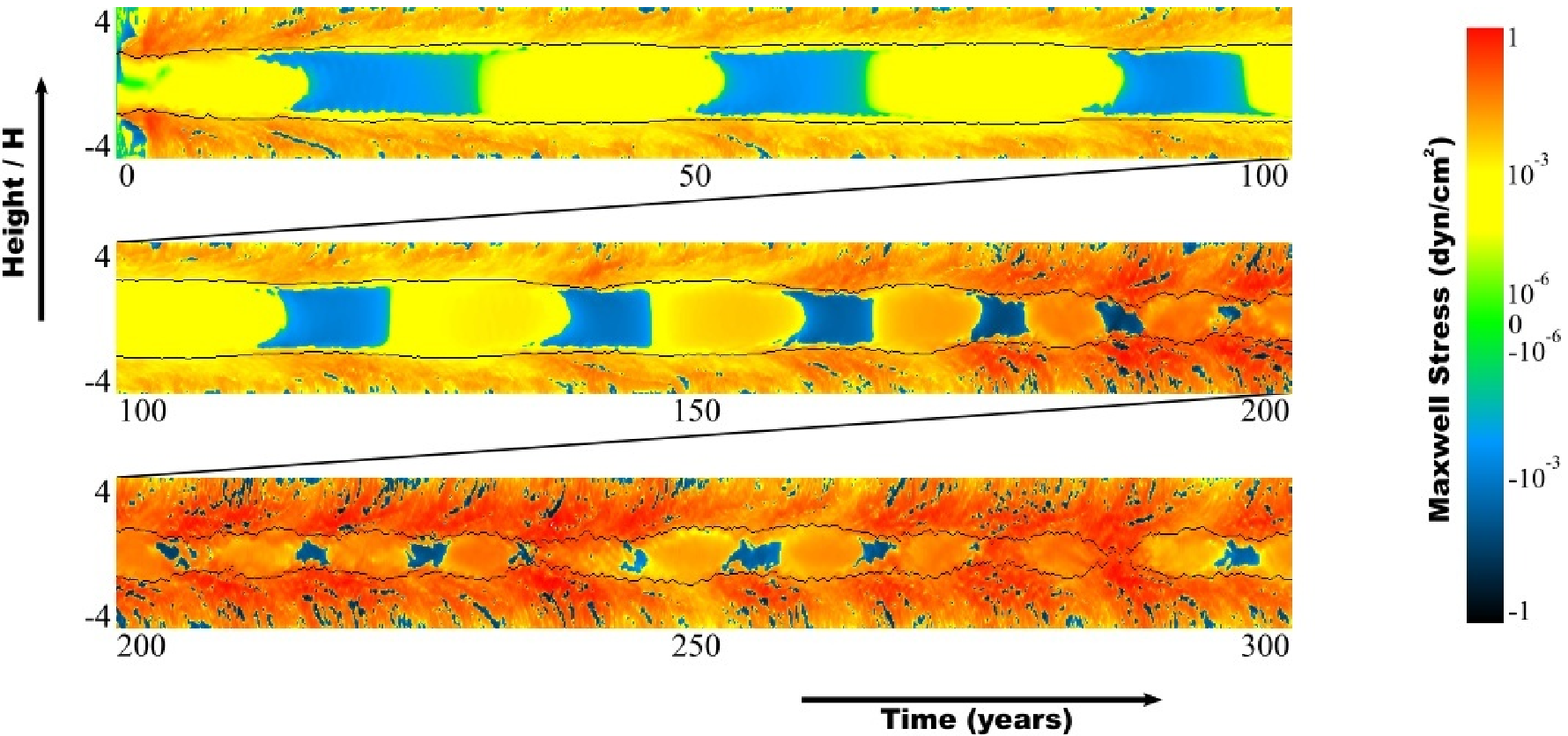}

  \figcaption{\sf Magnetic accretion stress versus height and time in
    a resistive MHD calculation of a patch of the protosolar disk at
    1~AU.  The stress is proportional to the radial and toroidal
    components of the magnetic field.  It is horizontally-averaged
    across the patch and is plotted using a double-logarithmic color
    scale, with outward angular momentum transport in red and yellow,
    and inward transport in black and blue.  The calculation lasts
    300~years and each panel shows a 100-year interval.  Black curves
    mark the edges of the undead zone where the resistivity suppresses
    magnetorotational turbulence.  Magnetic activity in the undead
    zone is driven by the diffusion of magnetic fields from the
    turbulent surface layers.  After 150~years, the activity grows
    stronger due to an increase in magnetic coupling as ionized gas is
    mixed from the surface layers toward the
    midplane.\label{fig:tzplot}}
\end{figure}

The two calculations give similar results over 130~orbits.  A net
radial magnetic field generated in the turbulent surface layers is
mixed to the undead zone where it diffuses to the midplane.  The shear
in the undead zone generates toroidal fields
(figure~\ref{fig:snapshot}) of the opposite sign, leading to magnetic
stresses that transfer angular momentum outward.  Lying just inside
the undead zone is a dissipation layer about $\frac{1}{2}H$ thick that
receives magnetic fields when turbulent motions overshoot the edge of
the well-coupled gas.  The resistivity is too high for
magneto-rotational turbulence and dissipates the tangled component of
the fields.  The Ohmic diffusion time $H^2/\eta$ at the base of the
dissipation layer is about 10~orbits, so that the layer acts as a
filter delivering to the undead zone a magnetic field that is a recent
time-average of the large-scale field at the edge of the turbulent
layer.  The radial field in the dissipation layer and at the midplane
reverses every 20-50~orbits.  The midplane toroidal field changes at
approximately the rate $-\frac{3}{2}\Omega B_r$ due to the shear,
indicating that local shear generation dominates other source terms
\citep{ts07}.  The shearing reverses the toroidal field about
10~orbits after the radial field, so that the field lines swing around
to trail the rotation and the magnetic stress again becomes positive
(figure~\ref{fig:tzplot}).  A related process occurs in the Sun, where
magnetic fields generated in the convection zone are carried down to
the tachocline and grow there through differential rotation
\citep{p93}.  The protostellar disk differs in that shear occurs
throughout the flow, and the turbulence is shut off by the resistivity
of the gas rather than a transition to radiative energy transport.

The resistivity at all heights in the disk is typically unchanged as
gas mixes between the surface layers and interior, because the
recombination on the grains is fast enough to keep the ionization near
its local equilibrium value.  However with the grains removed,
recombination is slow enough for ionized gas to reach the midplane
\citep{is05}.  In our calculation with no grains, the undead zone
shrinks after 150~years as ionized gas is carried down from the
surface layers.  The flow settles after 200~years into a new steady
state with an overall accretion rate 20~times greater.  The midplane
remains magneto-rotationally stable (figure~\ref{fig:tzplot}).

\section{EMPIRICAL CONSTRAINTS}

\begin{figure}[tb!]
  \epsscale{0.75}
  \plotone{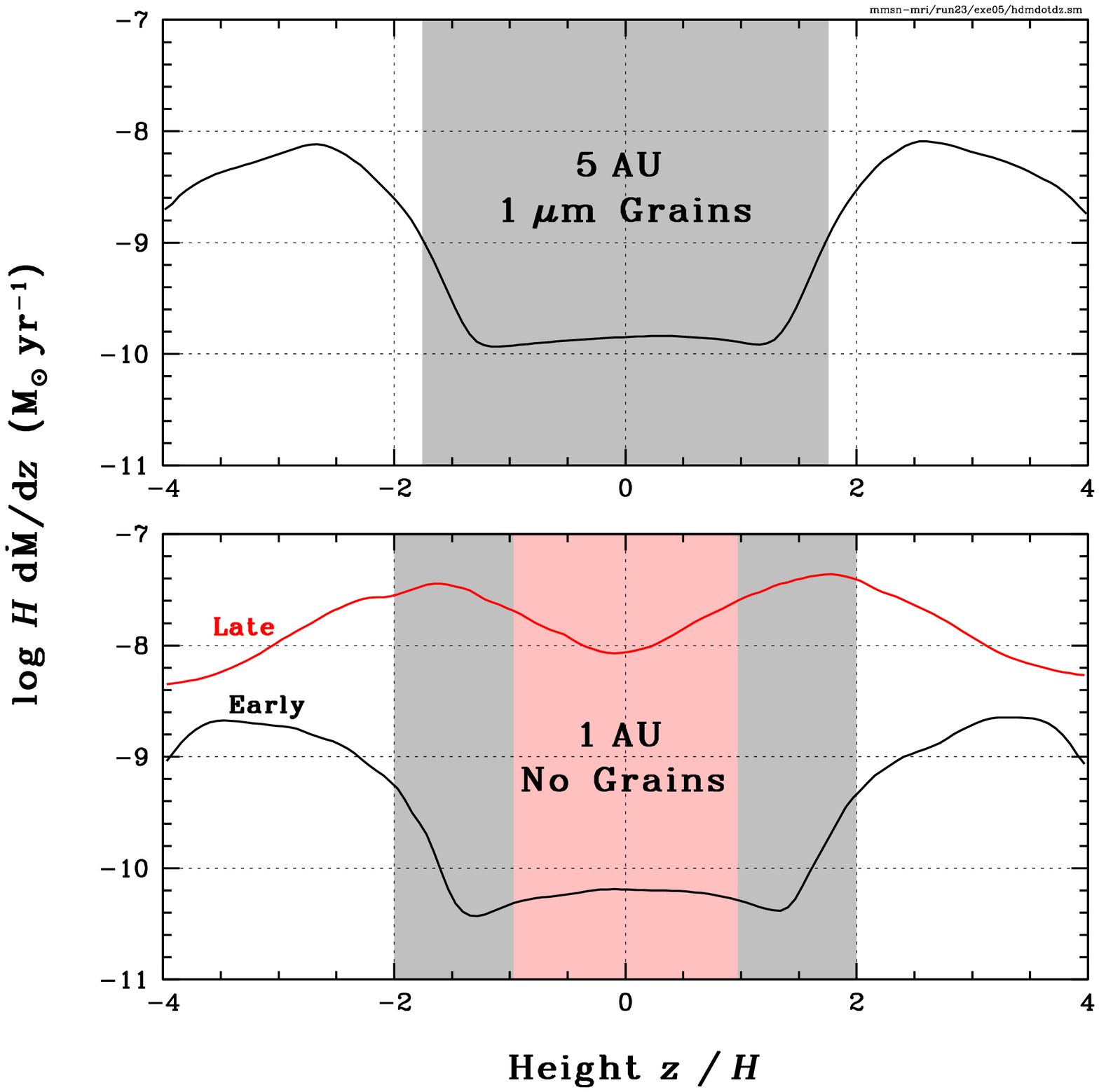}

  \figcaption{\sf Accretion rate corresponding to the total stress in
    the two MHD calculations located at 5~AU (top) and 1~AU (bottom).
    The accretion rate per density scale height $H$ is plotted against
    the distance $z$ from the midplane.  The results are averaged from
    20 to 120~orbits (solid black curves at top and bottom) and after
    mixing reduces the size of the undead zone, from 200 to 300~orbits
    (red curve in lower panel).  The vertically-integrated accretion
    rates are $2.2\times 10^{-8}$, $6.4\times 10^{-9}$ and $1.5\times
    10^{-7}$ Solar masses per year respectively.  Shading of the same
    colors marks the undead zone.\label{fig:hdmdotdz}}
\end{figure}

The mass flow rate corresponding to the stresses in the MHD
calculations (figure~\ref{fig:hdmdotdz}) is sufficient for the disk to
accrete on the star within a few million years, the lifetime inferred
from the disk fractions in young star clusters of different ages
\citep{hl01}.  The magnetic contribution to the stress is several
times greater than the hydrodynamic contribution in the dissipation
layer, and at times when the fields are strongest in the midplane.
The time-averaged midplane magnetic and hydrodynamic contributions are
comparable.  The undead zone rather than being inactive has a mean
mass flow rate 4\% to 61\% of that in the turbulent layers.  The
midplane gas moves toward the star on average, but flows away from the
star immediately after the reversals in the radial magnetic field when
the field lines lead in the rotation.  The root-mean-square midplane
field strength found in the calculations is about 0.1~Gauss at 5~AU
and 1~Gauss at 1~AU, in the range 0.1--7~Gauss inferred from the
relict magnetization of primitive meteorites that were last melted in
the asteroid belt during the formation of the Solar system
\citep{ch91}.  The agreement is consistent with the idea that the
planet formation region is a laminar, magnetized accretion flow for
some part of its history.

The original dead zone picture with a fixed mass column of accreting
material \citep{g96} yields a unique mass accretion rate approximately
$10^{-8}$ Solar masses per year, independent of the stellar mass.  In
contrast, observations of T~Tauri stars show the accretion rate
increases with the stellar mass and has a spread of about two decades
at Solar mass \citep{hd06}.  We suggest based on
figure~\ref{fig:hdmdotdz} that a similar range of accretion rates can
be produced by varying the magnetic flux and the abundance of small
dust grains.

The radial transport of material over large distances in the early
Solar system is indicated by the presence of crystalline silicate
grains in Comet 81P/Wild 2 \citep{bt06}.  The crystallization requires
temperatures of 1000~K that were found near the Sun, while much of the
comet is icy and formed at temperatures below 200~K in the outer Solar
system.  It remains to be seen whether grains can be carried large
radial distances in the turbulent surface layers of the protosolar
disk without sinking into the undead and dead zones.

\section{DISCUSSION}

We have found that magnetic stresses control the dead zone evolution
even with a full complement of micron-sized grains, in a minimum-mass
protosolar disk ionized only by stellar X-rays and radionuclide decay.
Consequently, smooth large-scale magnetic fields can occur in a wide
range of protostellar disks and drive flows in the dead zone before
and during planet formation as well as afterwards.

Models of the early evolution of the solids in protostellar disks
depend crucially on the dynamics of the gas and will be affected by
the presence of a laminar midplane accretion flow.  The gas drag
forces in turbulence cause collisions between the particles, leading
to the growth or destruction of solid bodies depending on the speeds
involved \citep{dd05}.  In laminar gas, grains settle toward the
midplane on timescales much less than the disk lifetime \citep{nn81}.
Layered accretion can cause the rapid formation of large bodies as
particles colliding in the turbulent surface layers grow large enough
to fall into the laminar interior where destructive collisions are
rare \citep{c07}.

Protoplanets migrate away from the locations where they first formed,
by exchanging orbital angular momentum with the disk gas through
gravitational torques \citep{gt79,w97}.  The rapid sunward migration
of the cores of Jupiter and Saturn is a difficulty for the standard
model of giant planet formation by core accretion \citep{hb05}.  The
migration rate depends on the distribution of gas near the planet,
which can be altered by magnetic forces \citep{t03}.  The midplane
magnetic pressure in our calculations reaches a few percent of the gas
pressure, sufficient for radial variations to halt or reverse the
orbital migration.  The outer edge of the undead zone is a plausible
location for radial magnetic gradients large enough to stop the
migration, and magnetic effects are a potential solution to the
problem of the rapid loss of planetary embryos through migration.

An important question for future studies is how the mass flow rate
varies with the distance from the star.  The fraction of the column
coupled to the magnetic fields generally increases with the radius.
Possibly the inflowing gas will pile up at some location, leading to
the development of a local radial pressure maximum \citep{kl07}.  If
so, gas drag forces push solid particles toward the maximum, providing
favorable conditions there for the growth of planetesimals
\citep{hb03}.  The further build-up of material could lead to
self-gravitational instability and an episode of rapid accretion on
the star \citep{g96,al01} similar to those associated with intervals
of strong mass outflow from young stellar objects \citep{r89}.

\acknowledgements

Part of this work was carried out at the Jet Propulsion Laboratory,
California Institute of Technology using the JPL Supercomputing
Facility and with support from the JPL Research and Technology
Development and NASA Solar Systems Origins Programs.


\begin{thebibliography}{}
\bibitem[Armitage et al.(2001)]{al01} Armitage P. J., Livio M. \&
  Pringle J. E. 2001, \mnras, 324, 705
\bibitem[Balbus \& Hawley(1991)]{bh91} Balbus S. A. \& Hawley
  J. F. 1991, \apj, 376, 214
\bibitem[Blaes \& Balbus(1994)]{bb94} Blaes O. M. \& Balbus
  S. A. 1994, \apj, 421, 163
\bibitem[Blandford \& Payne(1982)]{bp82} Blandford R. D. \& Payne
  D. G. 1982, \mnras, 199, 883
\bibitem[Brandenburg et al.(1995)]{bn95} Brandenburg A., Nordlund
  \AA., Stein R. F. \& Torkelsson U. 1995, \apj, 446, 741
\bibitem[Brownlee et al.(2006)]{bt06} Brownlee D. et al. 2006,
  Science, 314, 1711
\bibitem[Ciesla(2007)]{c07} Ciesla F. 2007, \apj, 654, L159
\bibitem[Cisowski \& Hood(1991)]{ch91} Cisowski S. M. \& Hood L. L.,
  in ``The Sun in Time'', ed. C. P. Sonett, M. S. Giampapa \&
  M. S. Matthews (Tucson: Univ. of Arizona Press), 761
\bibitem[Dullemond \& Dominik(2005)]{dd05} Dullemond C. P. \&
  Dominik C. 2005, \aap, 434, 971
\bibitem[Fleming \& Stone(2003)]{fs03} Fleming T. \& Stone J. M. 2003,
  \apj, 585, 908
\bibitem[Fromang et al.(2002)]{ft02} Fromang S., Terquem C. \& Balbus
  S. A. 2002, \mnras, 329, 18
\bibitem[Gammie(1996)]{g96} Gammie C. F. 1996, \apj, 457, 355
\bibitem[Garmire et al.(2000)]{gf00} Garmire G., et al. 2000, \aj, 120,
  1426
\bibitem[Goldreich \& Tremaine(1979)]{gt79} Goldreich P. \& Tremaine
  S. 1979, \apj, 233, 857
\bibitem[Haghighipour \& Boss(2003)]{hb03} Haghighipour N. \&
  Boss A. P. 2003, \apj, 583, 996
\bibitem[Haisch et al.(2001)]{hl01} Haisch K. E. Jr., Lada E. A. \&
  Lada C. J. 2001, \apj, 553, 153
\bibitem[Hartmann et al.(2006)]{hd06} Hartmann L., D'Alessio P.,
  Calvet N. \& Muzerolle J. 2006, \apj, 648, 484
\bibitem[Hawley et al.(1995)]{hg95} Hawley J. F., Gammie C. F. \&
  Balbus S. A. 1995, \apj, 440, 742
\bibitem[Hayashi et al.(1985)]{hn85} Hayashi C., Nakazawa K. \&
  Nakagawa Y., in ``Protostars and Planets II'', ed. D. C. Black \&
  M. S. Matthews (Tucson: Univ. of Arizona Press), 1100
\bibitem[Hubickyj et al.(2005)]{hb05} Hubickyj O., Bodenheimer P. \&
  Lissauer J. J. 2005, \icarus, 179, 415
\bibitem[Igea \& Glassgold(1999)]{ig99} Igea J. \& Glassgold
  A. E. 1999, \apj, 518, 848
\bibitem[Ilgner \& Nelson(2006)]{in06a} Ilgner M. \& Nelson
  R. P. 2006, \aap, 445, 205
\bibitem[Inutsuka \& Sano(2005)]{is05} Inutsuka S. \& Sano T. 2005,
  \apj, 628, L155
\bibitem[Kretke \& Lin(2007)]{kl07} Kretke K. A. \& Lin D. N. C. 2007,
  \apj, 664, L55
\bibitem[Miller \& Stone(2000)]{ms00} Miller K. A. \& Stone
  J. M. 2000, \apj, 534, 398
\bibitem[Nakagawa et al.(1981)]{nn81} Nakagawa Y., Nakazawa K. \&
  Hayashi C. 1981, \icarus, 45, 517
\bibitem[Parker(1993)]{p93} Parker E. N. 1993, \apj, 408, 707
\bibitem[Reipurth(1989)]{r89} Reipurth B. 1989, Nature, 340, 42
\bibitem[Sano et al.(2000)]{sm00} Sano T., Miyama S. M., Umebayashi
  T. \& Nakano T. 2000, \apj, 543, 486
\bibitem[Sano \& Stone(2002)]{ss02b} Sano T. \& Stone J. M. 2002,
  \apj, 577, 534
\bibitem[Stepinski(1992)]{s92} Stepinski T. F. 1992, \icarus, 97, 130
\bibitem[Stone \& Norman(1992)]{sn92b} Stone J. M. \& Norman
  M. L. 1992, \apj, 80, 791
\bibitem[Stone et al.(1996)]{sh96} Stone J. M., Hawley J. F., Gammie
  C. F. \& Balbus S. A. 1996, \apj, 463, 656
\bibitem[Terquem(2003)]{t03} Terquem C. E. J. M. L. J. 2003, \mnras,
  341, 1157
\bibitem[Turner et al.(2007)]{ts07} Turner N. J., Sano T. \&
  Dziourkevitch N. 2007, \apj, 659, 729
\bibitem[Ward(1997)]{w97} Ward W. R. 1997, \icarus, 126, 261
\bibitem[Wardle \& K\"onigl(1993)]{wk93} Wardle M. \& K\"onigl
  A. 1993, \apj, 410, 218
\bibitem[Wardle(2007)]{w07} Wardle M. 2007, \apss, 311, 35
\end{thebibliography}
\end{document}